\documentclass{pic2012}
\def\runauthor{Jakub C\'uth}
\def\shorttitle{Charge asymmetry in $t\bar t$ production on the ATLAS.}

\begin{document}
\title{Charge asymmetry in top quark pair production \\
in the di-lepton channel at the ATLAS experiment}

\author{Jakub C\'uth}

\address{Institute of Physics, Academy of Sciences of Czech Republic\\
Na Slovance 1999/2, 182 21 Praha 8, Czech Republic\\
on behalf of the ATLAS Collaboration \\
e-mail: Jakub.Cuth@cern.ch
}

\maketitle

\abstracts{
The charge asymmetry in top quark pair production in the di-lepton $t\bar t$ channel using the ATLAS experiment at Large Hadron Collider is presented. The asymmetry is studied in the distribution of absolute rapidity difference between top and anti-top. The reconstruction of the $t\bar t$ system is based on a leading order matrix element. The lepton charge asymmetry is also presented. All results are consistent with the Standard Model.}


\section{Introduction and definitions}
The measurement \cite{Cecil} of the charge asymmetry is an important test of Quantum Chromodynamics (QCD). The top quark, according to QCD, is produced in quark--anti-quark pairs. This production is symmetric under the charge conjugation at leading order.

The asymmetry originates from the interference between box and born diagrams, and between ISR and FSR at next-to-leading order (NLO) of QCD. The main production mechanism $gg\to t\bar t$ (80\%)  is symmetric, however other NLO processes $gq\to t\bar t q$, $q\bar{q}\to t\bar t g$ are not. More information about the asymmetry can be found in \cite{Yvone}.

The top charge asymmetry is defined as

\begin{equation}
    A_C^{t\bar t} = \frac{N(\Delta|y| > 0 ) - N(\Delta|y|< 0 )}{N(\Delta|y| > 0) + N(\Delta|y| < 0)} \;, 
    \label{eq:topChAsymm}
\end{equation}

where rapidity difference $\Delta|y| \equiv |y_t| - |y_{\bar t}|$ ($y_t$ and $y_{\bar t}$ are rapidities of the top quark and anti-quark respectively) and $N$ are the number of events fulfilling the condition in brackets. 

Because the analysis is performed in the di-lepton channel, it is possible to measure the lepton charge asymmetry. The lepton charge asymmetry is studied in the distribution of the lepton absolute pseudorapidity difference and is defined as

\begin{equation}
    A_C^{\ell\ell} = \frac{N(\Delta|\eta| > 0) - N(\Delta|\eta| < 0 )}{N(\Delta|\eta| > 0) + N(\Delta|\eta|<0)}\;, 
    \label{eq:lepChAsym}
\end{equation}

where pseudorapidity difference $\Delta|\eta| \equiv |\eta_{\ell^{+}}| - |\eta_{\ell^{-}}|$ ($\eta_{\ell^{+}}$ and $\eta_{\ell^{-}}$ are pseudorapidities of anti-lepton and lepton respectively) and $N$ are the number of events fulfilling the condition in brackets. Both top and lepton charge asymmetry are predicted to be of the order of half a percent (MC@NLO generator \cite{mcnlo}).

\section{Data and Monte Carlo}
This study was performed using $4.7\;\rm{fb}^{-1}$ of data from $pp$ collisions at $\sqrt{s} = 7\;\rm{TeV}$ recorded by the ATLAS detector \cite{ATLAS} in 2011. The standard ATLAS framework was used for Monte Carlo (MC) generation and  detector simulation (GEANT4). The $t\bar t$ signal was simulated using the MC@NLO generator \cite{mcnlo}.

The background processes were determined mostly by Monte Carlo simulations except for Multijets/W+jets background which was estimated on data using the Matrix Method. The following main selection criteria were used:

\begin{itemize}
    \item a primary vertex with at least five tracks and at least two jets with transversal momentum\footnote{The transversal momentum and energy are defined as $p_T \equiv p\cdot\sin\theta$ and  $E_T \equiv E\cdot\sin\theta$ respectively, where the polar angle $\theta$ is measured with respect to the LHC beamline.} $p_{T} > 25\,\rm{GeV}$ and rapidity $|\eta|<2.5$.
    \item events with exactly two leptons with opposite sign, with $E_T > 25\,\rm{GeV}$ for electrons and/or $p_{T} > 20\,\rm{GeV}$ for muons.
    \item di-lepton ($ee$ and $\mu\mu$) invariant mass more than $10\,\rm{GeV}$ away from the $Z$ boson mass 
    \item the magnitude of event missing transverse momentum $E_\mathrm{T}^\mathrm{miss}$ larger than $60\,\rm{GeV}$ for the $ee$ and $\mu\mu$ channels, and the sum of lepton and jet $E_T$ larger than $130\,\rm{GeV}$ for the $e\mu$ channel
\end{itemize}

\begin{table}
    \centering
    \begin{tabular}{l*{3}{r@{$\;\pm\;$}r}}
        Channel             & \multicolumn{2}{c}{$ee$}  & \multicolumn{2}{c}{$e\mu$}  & \multicolumn{2}{c}{$\mu\mu$}  \\
        \hline
        \hline
        $t\bar t$           &         590 & 60          &          4400 &  500        &          1640 &  170          \\
        $Z \to ee/\mu\mu$   &          19 &  7          &   \multicolumn{2}{c}{-}     &            83 &   29          \\
        $Z \to \tau\tau$    &          19 &  7          &           180 &   60        &            67 &   23          \\
        Single top          &          30 &  2          &           230 &   20        &            82 &    7          \\
        Dibosons            &           9 &  1          &            70 &    4        &            23 &    2          \\
        Multijets/ W+jets   &          70 & 36          &           250 &  130        &            32 &   17          \\
        Total               &         740 & 70          &          5100 &  500        &          1930 &  170          \\
        \hline
        Data                & \multicolumn{2}{c}{732}   &  \multicolumn{2}{c}{5305}   &  \multicolumn{2}{c}{2010}  
    \end{tabular}
    \caption{Expected and observed number of events \protect\cite{Cecil}.}
\end{table}

\section{Event reconstruction}
The input reconstruction objects are: two leptons (electron or muon), two jets and missing transverse energy. Because of the limited detector resolution, and under-constrained system with two neutrinos, the studied system is solved in many phase space points and for each solution the weight is calculated. This weight  is based on computing a probability distribution using the $gg\to t\bar t$ leading-order matrix element $\mathcal{M}(y)$ \cite{Cecil}. 
            
The weights are defined as:
\begin{equation}
w \equiv \frac{(2\pi)^4}{\varepsilon_1\varepsilon_2s}d\varepsilon_1 d\varepsilon_2 f_{PDF}(\varepsilon_1) f_{PDF}(\varepsilon_2) |\mathcal{M}(y)|^2 W(x,y) {\rm d}\Phi_n \;,
\label{eq:weigth}
\end{equation}

where $y$ are the assumed partonic kinematic values, $x$ are the reconstructed kinematic values, $\varepsilon_{1,2}$ are the gluon momentum fractions in the protons, $f_{PDF}$ is the parton density function,  $W(x,y)$ relates the partonic and reconstructed quantities and ${\rm d}\Phi_n$ is the phase space element. The reconstructed values of the top quarks rapidity were computed as the weighted average of all phase space points.

\begin{figure}[!thb]
    \begin{center}
        \includegraphics[width=.32\textwidth]{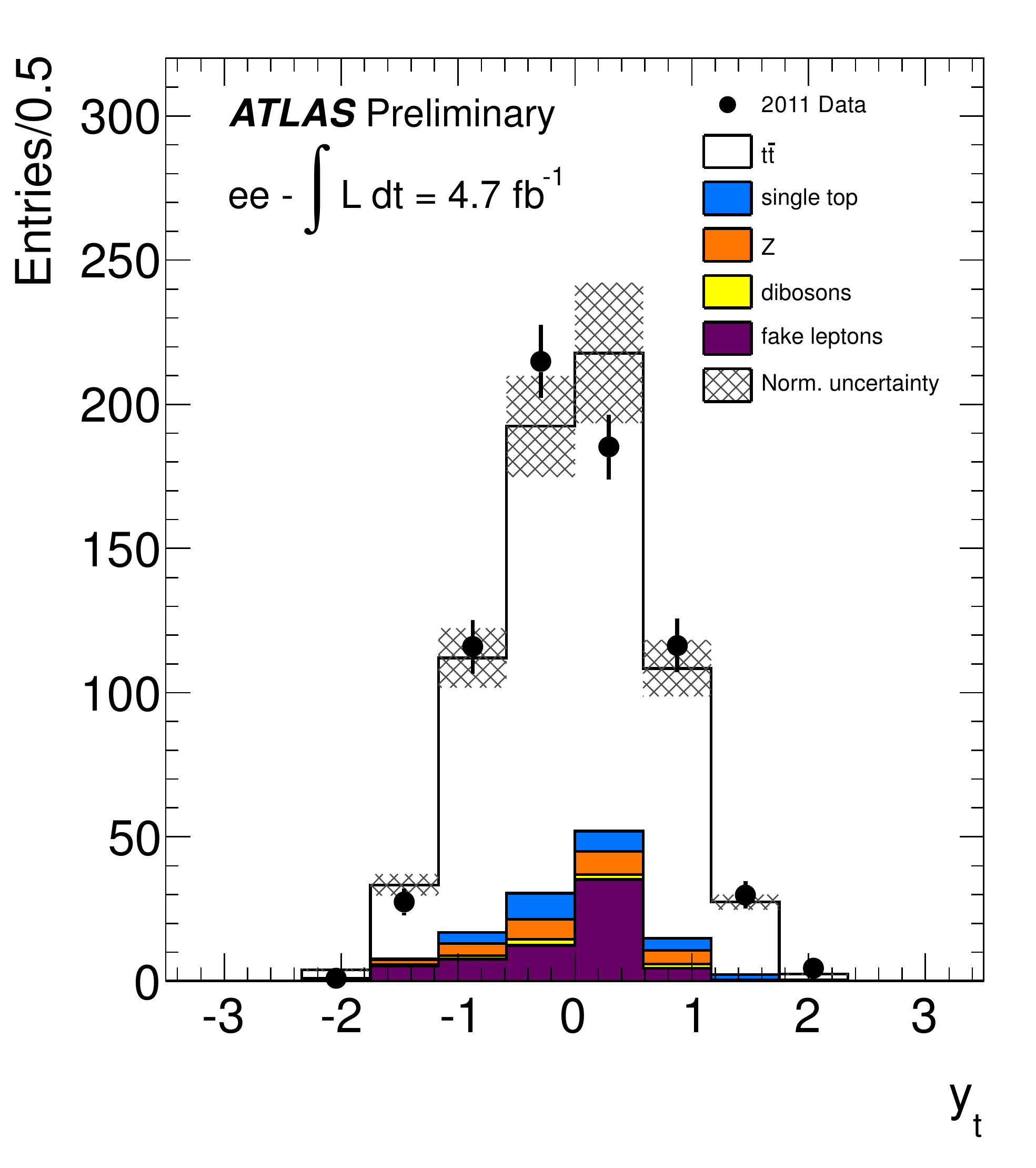}
        \includegraphics[width=.32\textwidth]{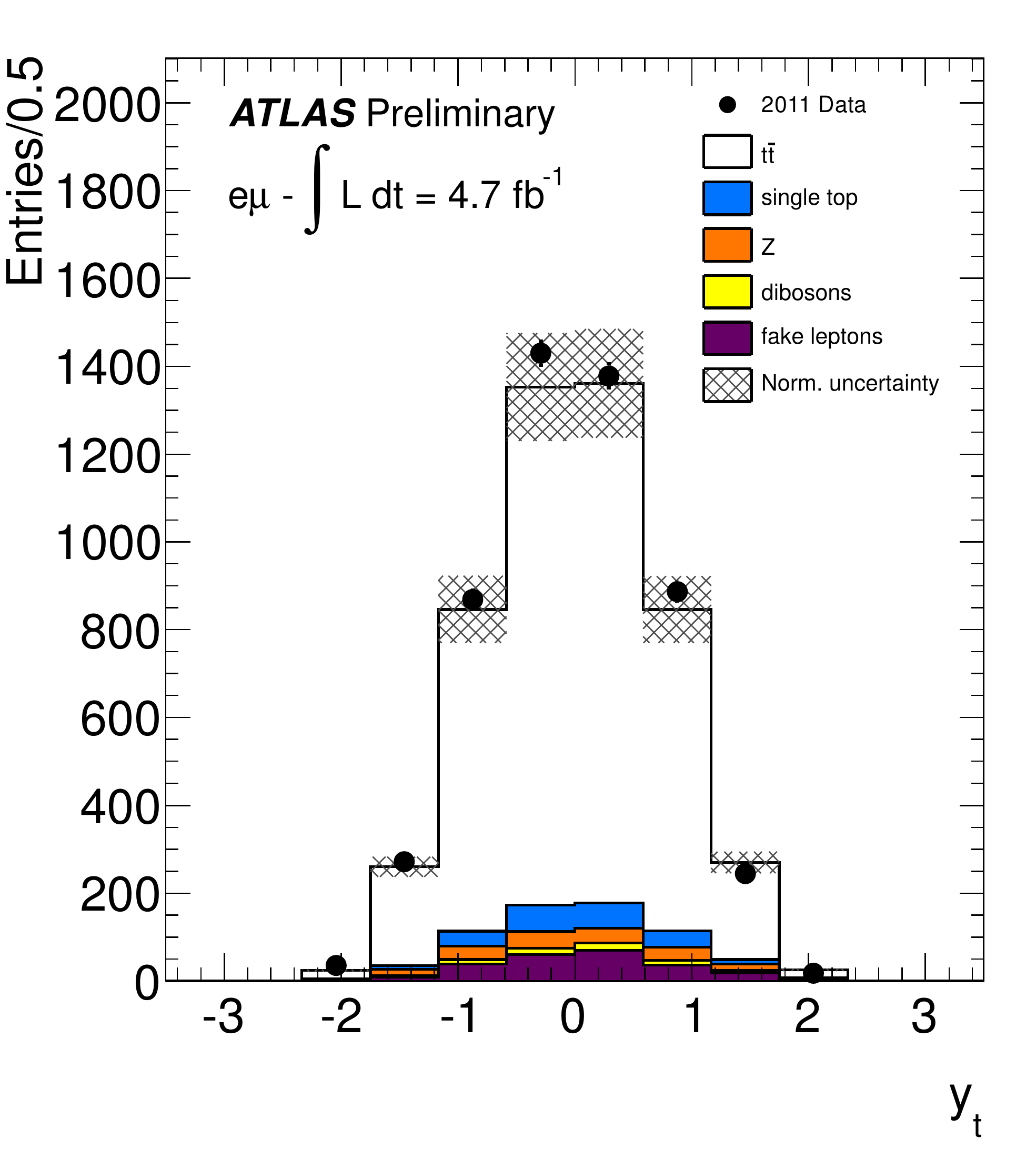}
        \includegraphics[width=.32\textwidth]{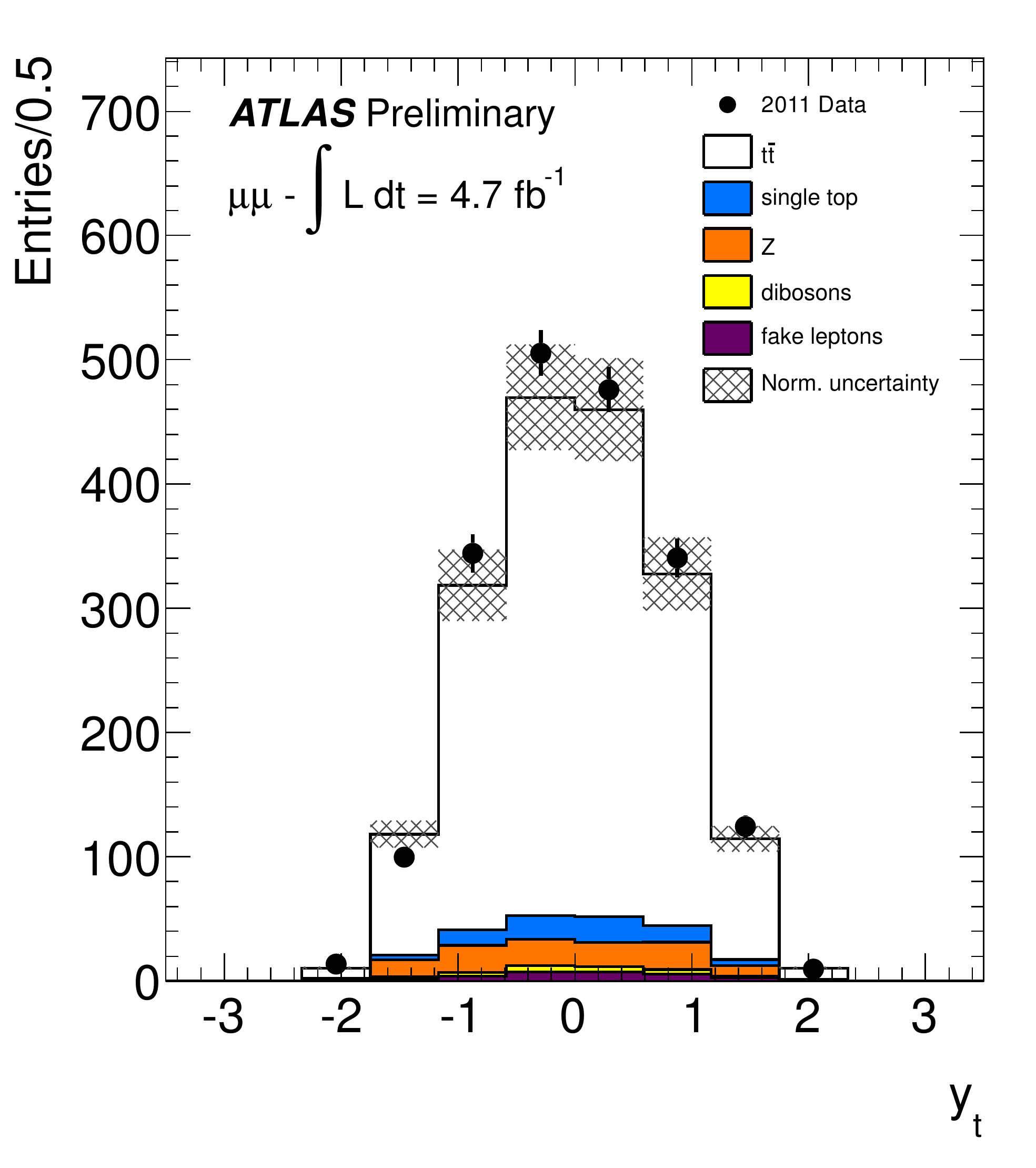}
        \caption{Measured and expected rapidity distribution of top and antitop quarks \protect\cite{Cecil}. The sub-figures show each channel in order from left to right: $ee$, $e\mu$ and $\mu\mu$. }
    \end{center}
\end{figure}

\section{Measurement}
 The background subtracted data asymmetries were corrected for distortion due to detector effects and event selection. These corrections were estimated from simulated $t \bar t$ sample with injected asymmetries. 
 
\begin{figure}[!thb]
    \begin{center}
        \includegraphics[width=.32\textwidth]{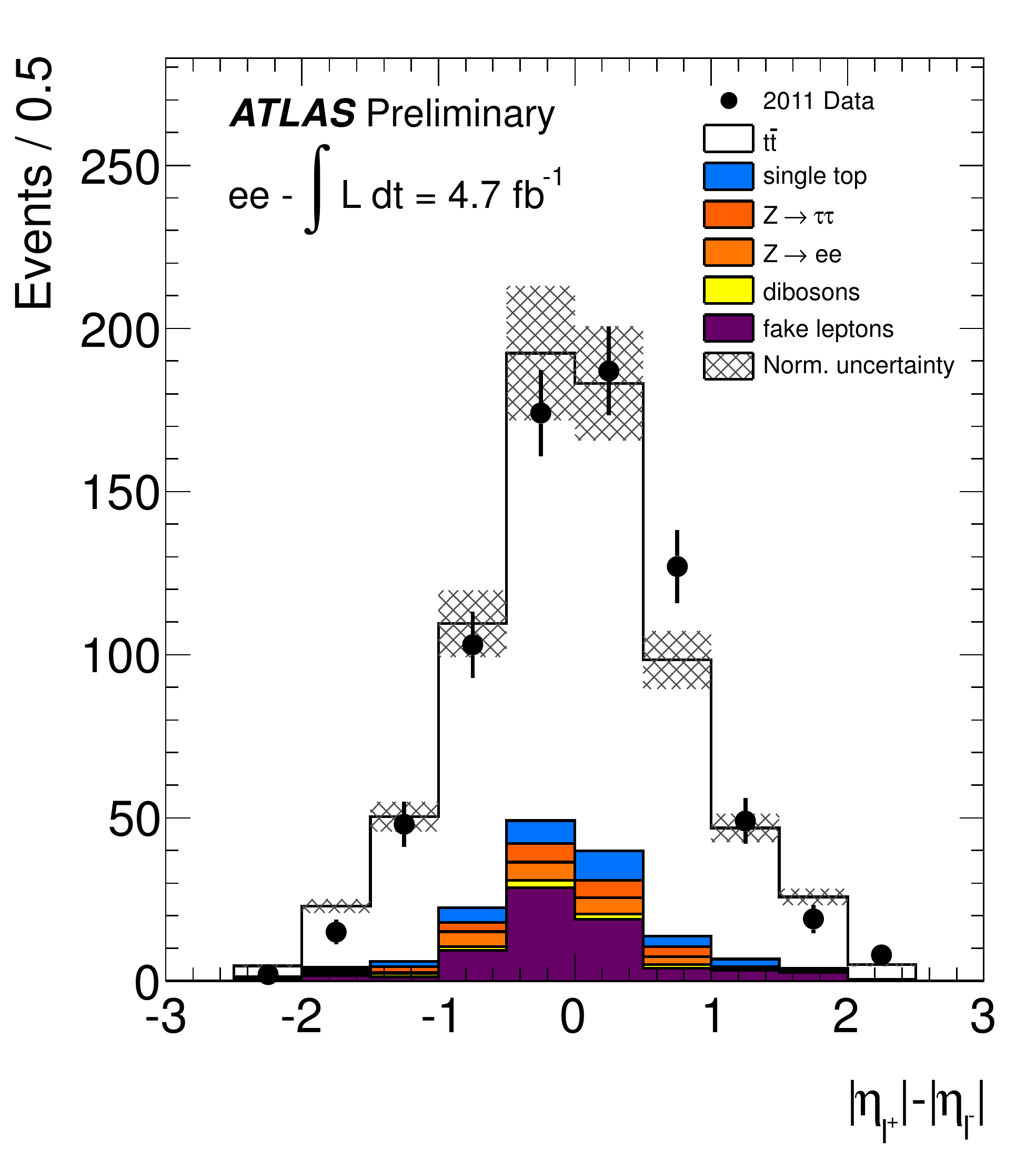}
        \includegraphics[width=.32\textwidth]{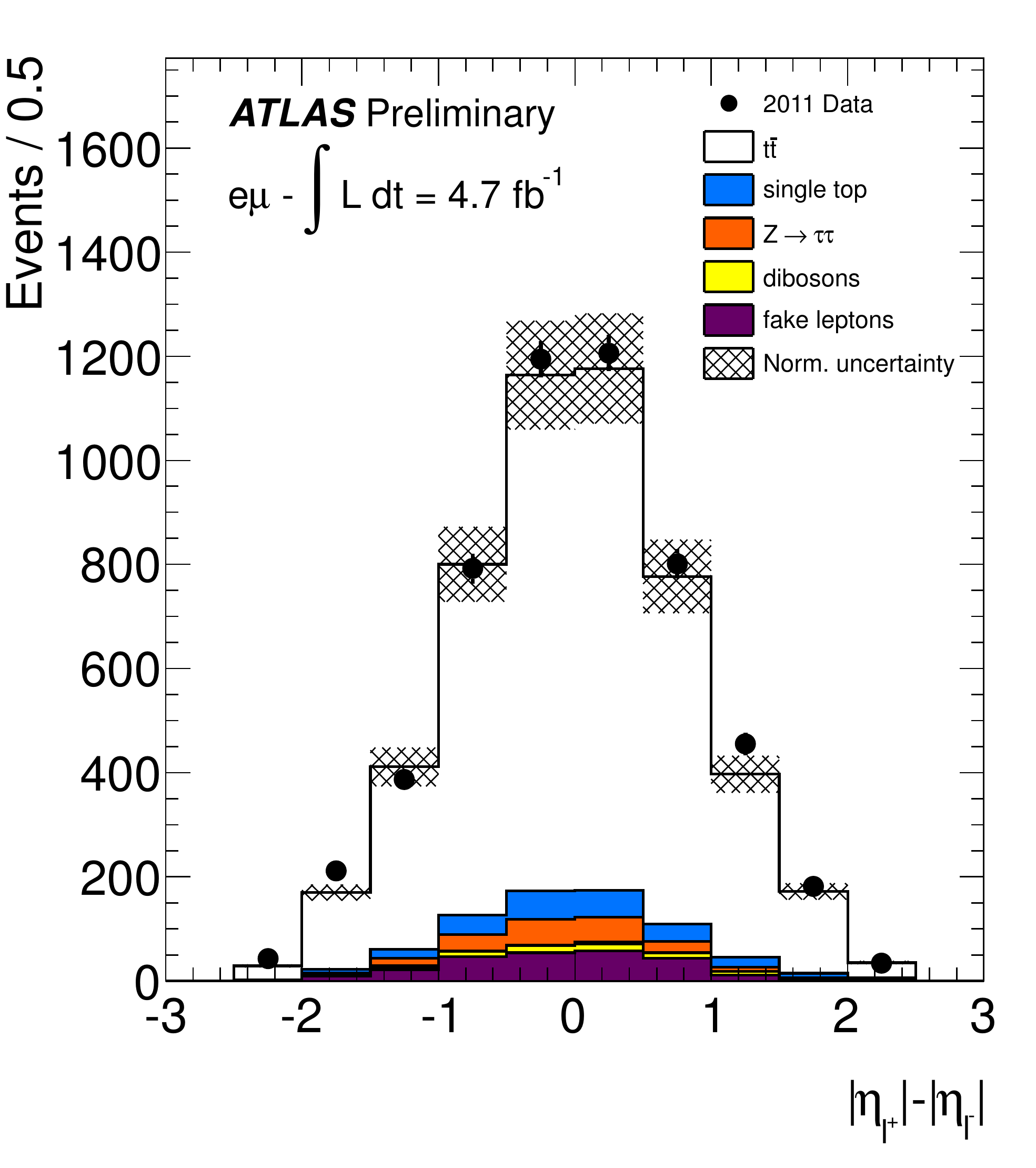}
        \includegraphics[width=.32\textwidth]{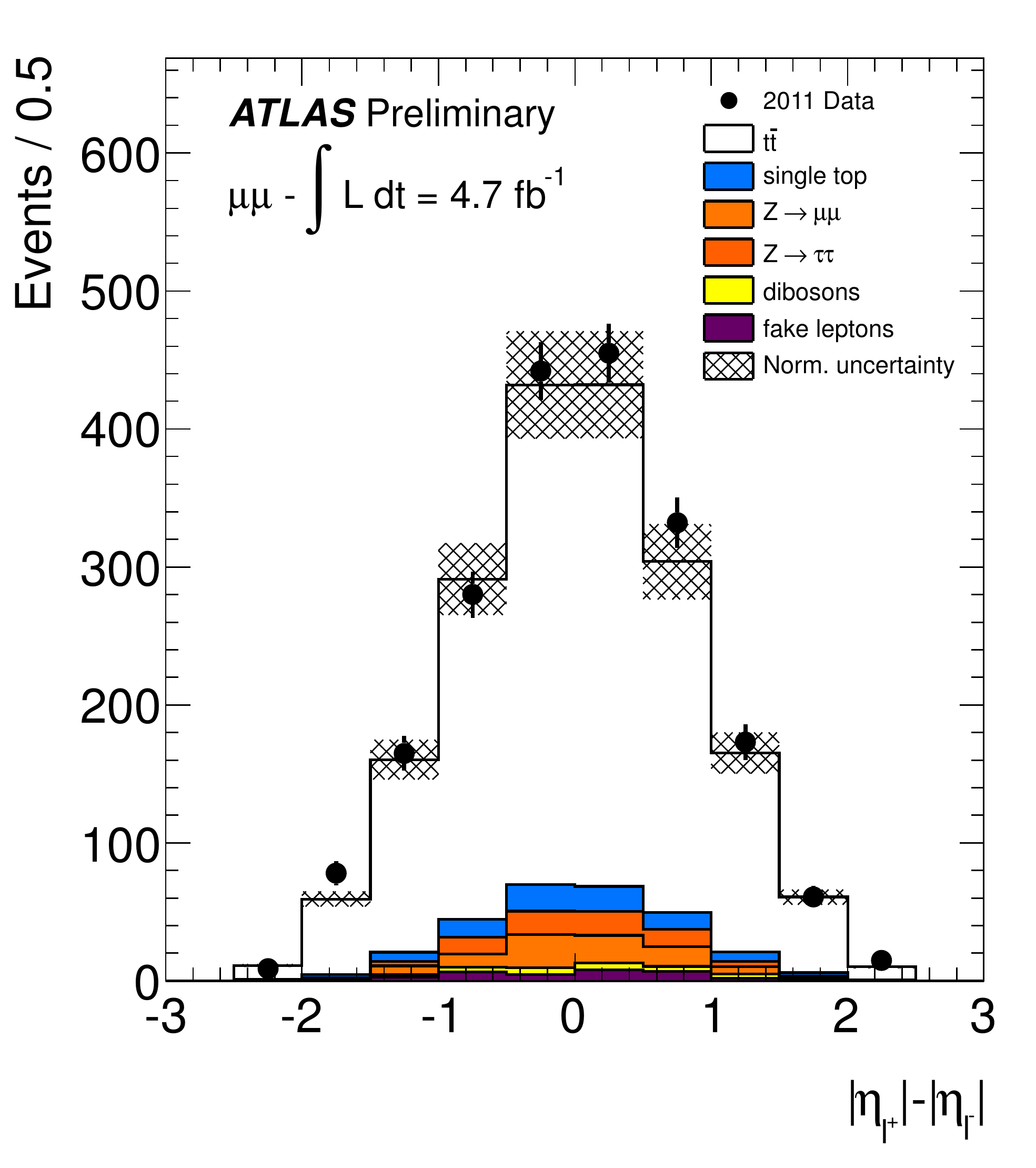}
        \caption{Measured and expected distribution of absolute pseudorapidity difference between positive and negative lepton \protect\cite{Cecil}. The sub-figures show each channel in order from left to right:  $ee$, $e\mu$ and $\mu\mu$. }
    \end{center}
\end{figure}

 The lepton charge asymmetry results, after subtraction of background and correction for detector acceptance and resolution, are

\vskip1mm
\begin{center}
    \begin{tabular}{l@{$\;=\;$}r@{$\;\pm\;$}r@{$\;$(stat.)$\;\pm\;$}r@{$\;$(syst.)}c}
        ${A_C^{\ell\ell}}$  &  0.091  & 0.041 & 0.029 & (${ee}$ channel),       \\
        ${A_C^{\ell\ell}}$  &  0.018  & 0.014 & 0.009 & (${e\mu}$ channel),     \\
        ${A_C^{\ell\ell}}$  &  0.026  & 0.023 & 0.009 & (${\mu\mu}$ channel).
    \end{tabular}
\end{center}
\vskip1mm

\begin{figure}[!thb]
    \begin{center}
        \includegraphics[width=.32\textwidth]{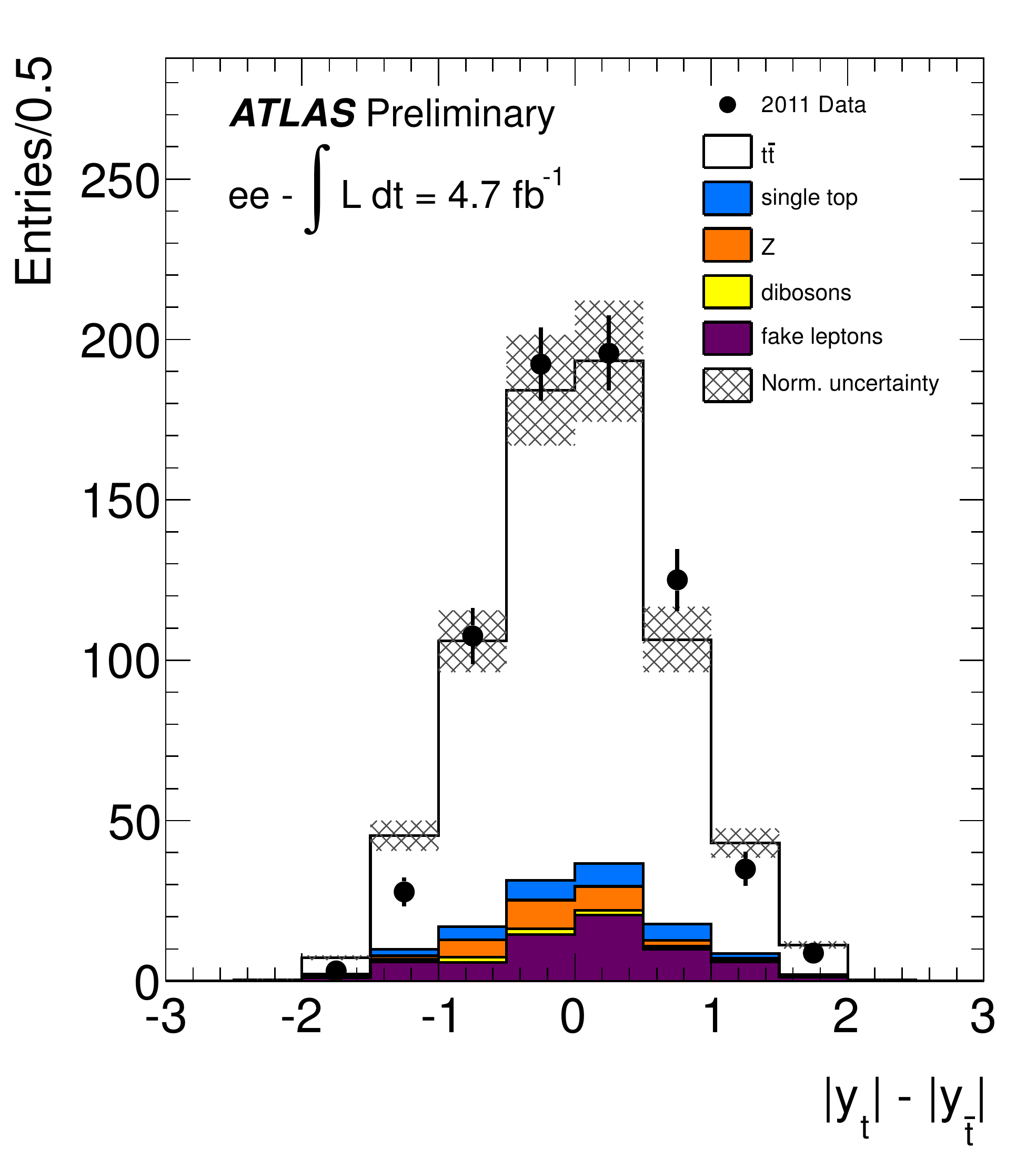}
        \includegraphics[width=.32\textwidth]{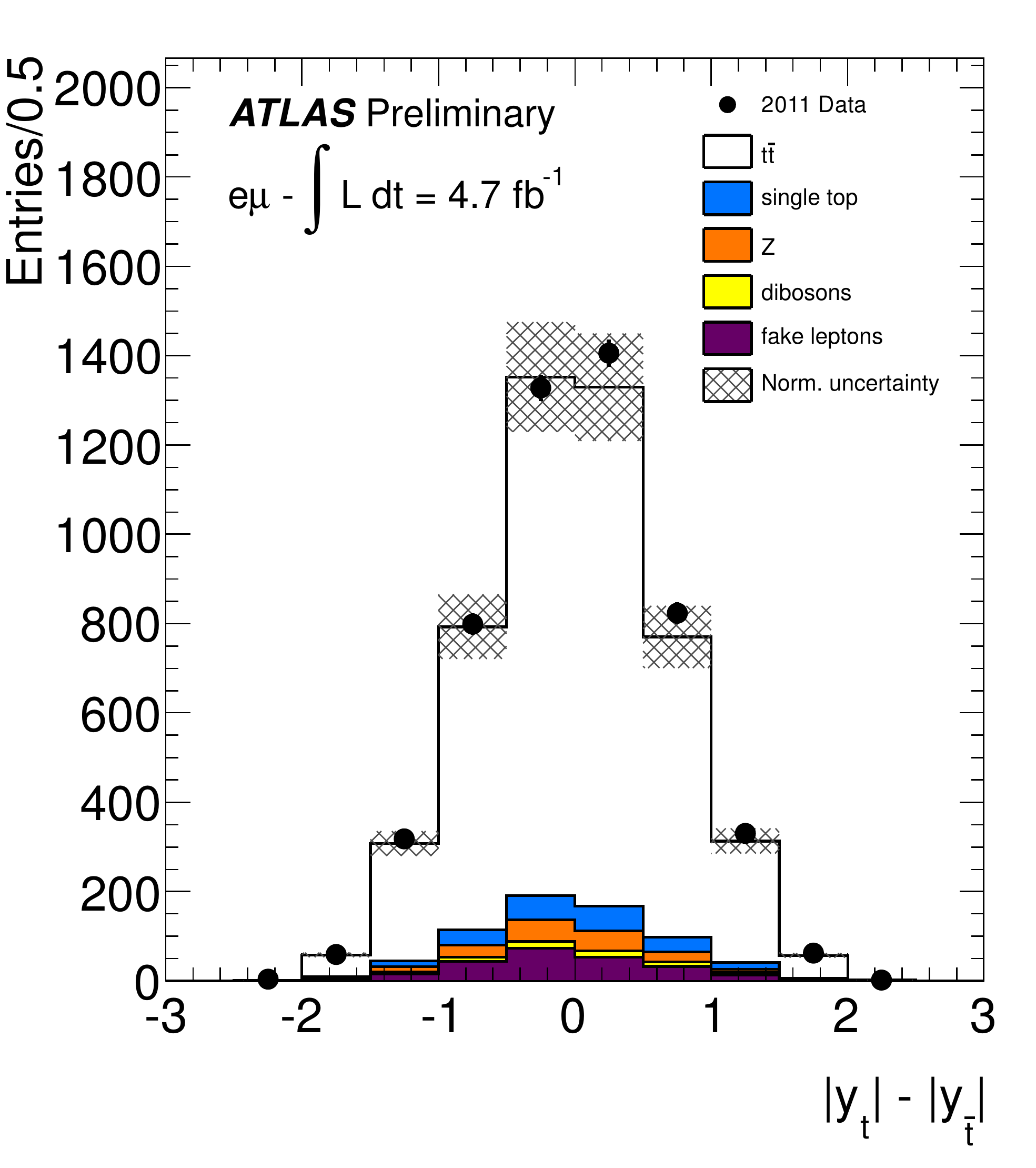}
        \includegraphics[width=.32\textwidth]{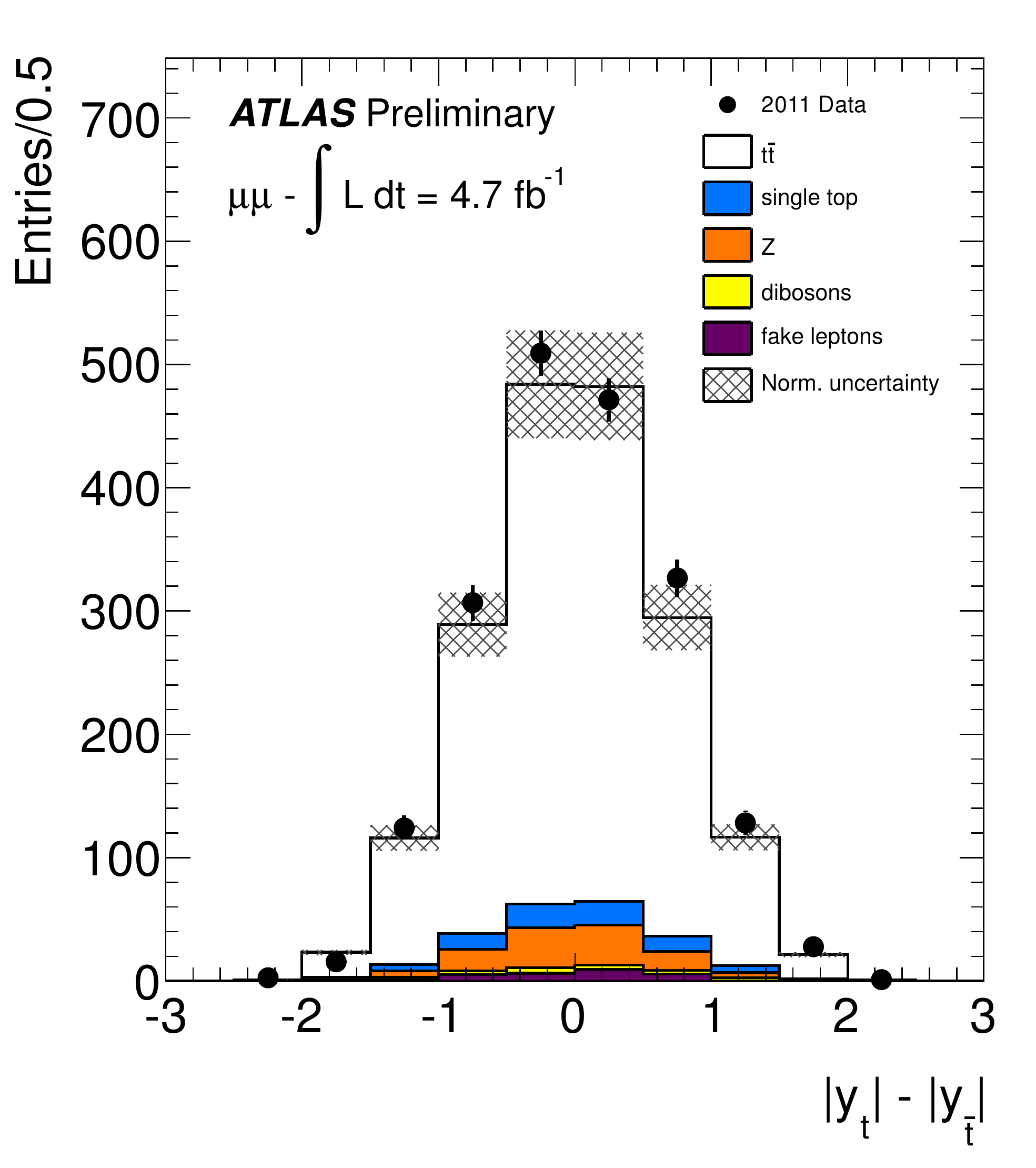}
        \caption{Measured and expected distribution of absolute rapidity difference between top and antitop \protect\cite{Cecil}. The sub-figures show each channel in order from left to right:  $ee$, $e\mu$ and $\mu\mu$.}
    \end{center}
\end{figure}

And the $t\bar t$ charge asymmetry is found to be:

\vskip1mm
\begin{center}
    \begin{tabular}{l@{$\;=\;$}r@{$\;\pm\;$}r@{$\;$(stat.)$\;\pm\;$}r@{$\;$(syst.)}c}
        $A_C^{t\bar t}$  &  0.079  & 0.087 & 0.028 & (${ee}$ channel),       \\
        $A_C^{t\bar t}$  &  0.078  & 0.029 & 0.017 & (${e\mu}$ channel),     \\
        $A_C^{t\bar t}$  &  0.000  & 0.046 & 0.021 & (${\mu\mu}$ channel).
    \end{tabular}
\end{center}
\vskip1mm

\section{Summary and Conclusion}
    The results presented here are from the first measurement of the top charge asymmetry in the di-lepton channel at LHC \cite{Cecil}.

    All channels were combined using the best linear unbiased estimator method (BLUE) and the final lepton charge asymmetry is

    \begin{center}
        $A_C^{\ell\ell}$ = 0.023 $\pm$ 0.012 (stat.) $\pm$ 0.008 (syst.).
    \end{center}

    The results for the top charge asymmetry were combined using BLUE and the result is
    \begin{center}
        $A_C^{t\bar t}$ = 0.057 $\pm$ 0.024 (stat.) $\pm$ 0.015 (syst.).
    \end{center}

    The top-based asymmetry was also studied in the lepton+jet channel \cite{LepAndJet} and results were combined with the measurement presented here to give

    \begin{center}
        $A_C^{t\bar t}$ = 0.029 $\pm$ 0.018 (stat.) $\pm$ 0.014 (syst.).
    \end{center}

    The current results are compatible with predictions of the Standard Model. Comparison of the results from other experiments can be found in another contribution of these proceedings \cite{Yvone}.

\section*{Acknowledgements} 
I would like to express my gratitude to my supervisor Roman Lys\'ak and other people from the top physics group for guidance and help with poster presentation.

\end{document}